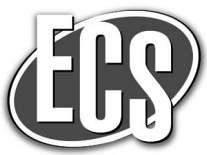

# Design of Bi-Tortuous, Anisotropic Graphite Anodes for Fast Ion-Transport in Li-Ion Batteries

V. Pavan Nemani,[a] Stephen J. Harris,[b,*] and Kyle C. Smith[a,c,*,z]

[a]*Department of Mechanical Science and Engineering, University of Illinois at Urbana-Champaign, Urbana, Illinois 61802, USA*
[b]*Materials Science Division, Lawrence Berkeley National Laboratory, Berkeley, California 94720, USA*
[c]*Computational Science and Engineering, University of Illinois at Urbana-Champaign, Urbana, Illinois 61802, USA*

Thick Li-ion battery electrodes with high ion transport rates could enable batteries that cost less and that have higher gravimetric and volumetric energy density, because they require fewer inactive cell-components. Finding ways to increase ion transport rates in thick electrodes would be especially valuable for electrodes made with graphite platelets, which have been shown to have tortuosities in the thru-plane direction about 3 times higher than in the in-plane direction. Here, we predict that bi-tortuous electrode structures (containing electrolyte-filled macro-pores embedded in micro-porous graphite) can enhance ion transport and can achieve double the discharge capacity compared to an unstructured electrode at the same average porosity. We introduce a new two-dimensional version of porous-electrode theory with anisotropic ion transport to investigate these effects and to interpret the mechanisms by which performance enhancements arise. From this analysis we determine criteria for the design of bi-tortuous graphite anodes, including the particular volume fraction of macro-pores that maximizes discharge capacity (approximately 20 vol.%) and a threshold spacing interval (half the electrode's thickness) below which only marginal enhancement in discharge capacity is obtained. We also report the sensitivity of performance with respect to cycling rate, electrode thickness, and average porosity/electroactive-material loading.
© The Author(s) 2015. Published by ECS. This is an open access article distributed under the terms of the Creative Commons Attribution 4.0 License (CC BY, http://creativecommons.org/licenses/by/4.0/), which permits unrestricted reuse of the work in any medium, provided the original work is properly cited. [DOI: 10.1149/2.0151508jes] All rights reserved.

Present-day oil demand and supply statistics show a clear need for alternatives in energy production, management, and storage. Amongst the various energy-storage devices available, Li-ion batteries have benefits of high gravimetric and volumetric energy-density.[1–3] Modern Li-ion battery electrodes are porous composites of solid-state active-material particles bound together by a conductive carbon-binder mixture, with an ion-conducting liquid electrolyte filling the pores. When a battery operates, electrons and ions are simultaneously transported to the surfaces of active-material particles, where electrochemical reactions take place. The rates at which ions are transported depend on the microscopic structure of the composite electrodes through a parameter called tortuosity. The microstructure in an electrode results from the particular choice of material constituents and processes that are used to fabricate the electrodes. To maximize the energy density of a battery, we would like to have electrodes with low porosity (maximizing the density of active material) and high thickness, reducing the number of inactive components (separators, current collectors) that are required for a given amount of active material, saving considerable cost. Unfortunately, electrodes with low porosity generally have high tortuosity,[4] making ion transport slow, an effect whose importance is magnified when electrodes are thick. Thus, techniques that produce thick, dense electrodes with enhanced ion transport could enable the development of batteries with high energy-density *and* high power density at a lower cost than is currently available.[5]

It has recently been shown that the morphology of active particles and the processing conditions used affect the electrode microstructure. For example, electrodes composed of plate- or flake-shaped particles of LiFePO$_4$ (Ref. [6]) and graphite (Ref. [7]) exhibit ordered orientation (self-assembly) that affects the ability to transport Li ions when they are confined into a dense state. Also, it has been shown that the tortuosity itself can be heterogeneous,[8] in addition to being anisotropic for electrodes comprised of non-spherical particles.[7]

Novel approaches have been developed recently to enhance ion-transport in Li-ion batteries by controlling electrode microstructure. Chiang and co-workers[9] used a co-extrusion process to manufacture half-cells containing low-tortuosity macro-pores embedded in sintered, micro-porous LiCoO$_2$. Here, structured electrodes with tuned dimensions achieved greater capacity than homogeneous electrodes with the same average porosity.[9] The mechanism for this enhancement was attributed to the improved Li-ion transport provided by the low-tortuosity macro-pores. Similarly Lu and Harris suggested that electrodes made with self-assembled particles could have low tortuosities in all direction.[30]

High-fidelity models of electrochemical transport are needed to guide the design of complex, multi-scale electrodes. In Ref. [9] scaling analysis of 'effective' tortuosity was used to explain experimentally observed enhancements in capacity. Subsequently, Cobb and Blanco[10] analyzed half-cells containing similar bi-tortuous structures using porous-electrode theory. They found that co-extruded macro-pore dimensions on order of 25 µm yielded the highest utilization of electroactive material for electrode thicknesses in the range of 150–300 µm.[10]

In the present work we predict the enhancement in cycling performance that could be achieved by structuring graphite anodes with low-tortuosity macro-pores. We perform these calculations on a full-cell in which the graphite anode is paired with a LiCoO$_2$ cathode. Our novel approach explicitly incorporates the anisotropy of ion transport in both electrodes using two-dimensional porous-electrode theory. With these predictions we show that the available capacity of electrodes with multi-scale porosity depends strongly on (1) the spacing between macro-pores in the anode and (2) the volume of the electrode comprised by the macro-pores. The analysis provides guidelines for the optimal design of macro-pores in graphite anodes.

The paper is organized as follows. We first describe the methods and materials used in the present simulations. Subsequently, results are presented to illustrate the time- and space-dependent electrochemical processes that occur in homogeneous and bi-tortuous electrodes. The dependence of performance on key design parameters for bi-tortuous electrodes is then explored, and the sensitivity of performance of optimized bi-tortuous electrodes to manufacturing and operational parameters is quantified. Subsequently, we conclude with a brief summary and highlight the implications of the present results in the broader context of Li-ion battery manufacturing and electrochemical energy-storage.

## Materials and Methods

*Bi-tortuous electrodes: Porosity and electroactive-material distribution.*— A full-cell containing a LiCoO$_2$ cathode and a graphite anode is simulated, as shown in Fig. 1a. Each of these electrodes is modeled as a two-dimensional continuum using porous-electrode theory (described later). A heterogeneous distribution of electroactive material and electrolyte is incorporated with spatially dependent electroactive-material volume-fraction $v_s(x, y)$ and micro-porosity $\varepsilon(x, y)$. Presently, we neglect the volume of conductive-carbon binder, in which case the local porosity and electroactive-material volume fraction are related by $v_s + \varepsilon = 1$. As shown in previous work,[7] the orientation of particles inside heterogeneous, composite electrodes affects the ion transport rates. We assume that graphite and LiCoO$_2$ particles orient parallel to the respective current collectors on which they are cast and calendered. Accordingly, in Fig. 1a horizontal-line pattern is used to represent the orientation of graphite platelets in the anode, while a cross-hatch pattern is used for oblate LiCoO$_2$ particles in the cathode (because electrodes containing oblate LiCoO$_2$ particles produce more isotropic microstructures than pressed graphite platelets[7]).

*Electrochemical Society Active Member.
[z]E-mail: kcsmith@illinois.edu

Here, we focus on particular designs for the anode that incorporate rectangular "macro-pores" (i.e., grooves when projected in the third dimension) that contain 100% electrolyte. These electrode structures are bi-tortuous in the sense that macro-pores possess a unit tortuosity, while the region containing electroactive material (i.e., where $v_s > 0$) exhibits a different,[9] direction-dependent[7] tortuosity. Figures 1b and 1c contrast the structures simulated for a homogeneous graphite anode with a bi-tortuous one. These macro-pores are assigned width $g$ and are separated by regular spacing $s$. The regular spacing between macro-pores enables the simulation of an infinite electrode by a single "unit cell" of a periodic structure.

We introduce constraints on the distribution of electroactive-material loading and porosity, such that the area-specific capacity is the same when comparing between bi-tortuous and homogeneous electrodes. In other words, local micro-porosity $\varepsilon$ (inside regions of the electrode containing electroactive material) is lower than the average porosity $\bar{\varepsilon}$. The extent of variation between these local and average values depends on the macro-pore coverage $v_{mp}$, which is defined as the volume fraction of the electrode covered by the macro-pore. For the two-dimensional electrode here macro-pore coverage is affected by the width of and spacing between macro-pores, $v_{mp} = g/s$ (a different expression can be derived for arbitrarily shaped and spaced macro-pores). Assuming negligible volume occupied by the conductive carbon-binder domains, the local micro-porosity depends on average porosity and macro-pore coverage in the following way:

$$\varepsilon = \frac{\bar{\varepsilon} - v_{mp}}{1 - v_{mp}}. \quad [1]$$

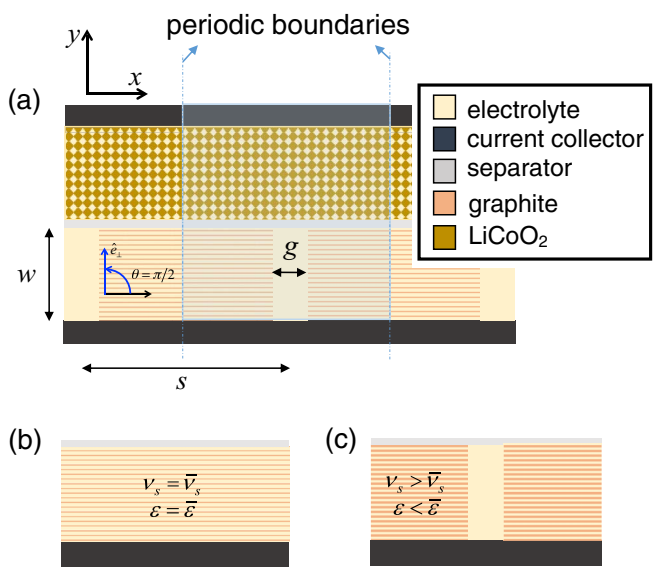

**Figure 1.** Schematic of material distribution in the two-dimensional full-cell simulated here. (a) Periodic boundaries (marked in blue) are used to simulate an infinite cell. The structured anode contains electrolyte-rich macro-pores with width $g$ and spacing $s$. Comparison of active-material volume-fraction and porosity for (b) a homogeneous graphite-anode and (c) bi-tortuous graphite anode.

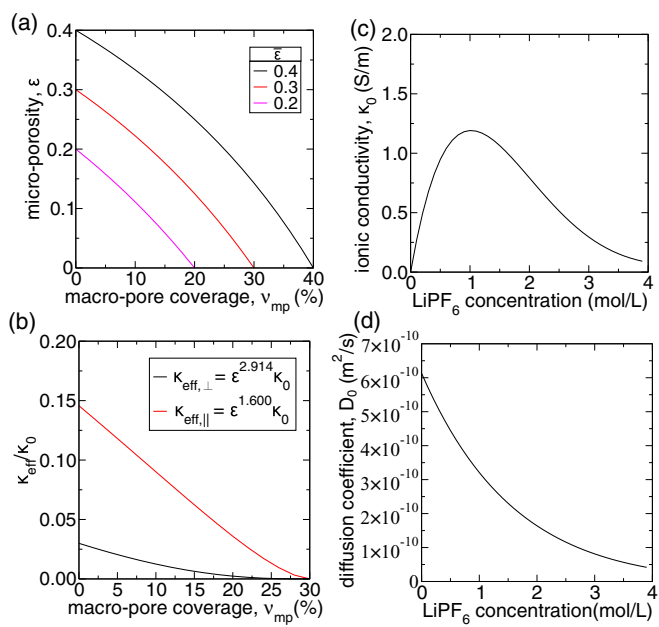

**Figure 2.** (a) Variation of micro-porosity $\varepsilon$ with increasing macro-pore coverage $v_{mp}$ for various average porosity levels $\bar{\varepsilon}$. (b) Variation of ionic-conductivity-tensor components with macro-pore coverage in the graphite anode (30% average porosity) along the directions perpendicular and parallel to the current collector. (c) Variation of bulk ionic-conductivity $\kappa_0$ and (d) bulk chemical-diffusion coefficient $D_0$ as a function of LiPF$_6$ concentration.[13]

The variation of local micro-porosity with macro-pore coverage is shown in Fig. 2a. These curves reveal that the maximum possible macro-pore coverage that can be achieved (i.e., where porosity vanishes) depends on the magnitude of average porosity (and implicitly on the average volume-fraction of electroactive material because $\bar{v}_s = 1 - \bar{\varepsilon}$). The local volume fraction of electroactive material $v_s$ (also called "loading") depends on the average value $\bar{v}_s$ and macro-pore coverage, as well:

$$v_s = \frac{\bar{v}_s}{1 - v_{mp}}. \quad [2]$$

*Anisotropic ion-transport parameters.—* Preferential orientation of pores in the microstructure of calendered graphite electrodes has been shown to produce anisotropic tortuosity that is high in the direction perpendicular to the current collector and low parallel to the current collector.[7] For the multi-dimensional system being modeled presently, this tortuosity anisotropy manifests in the effective diffusivity $\underline{\underline{D}}_{eff}$ and ionic-conductivity $\underline{\underline{\kappa}}_{eff}$ tensors, which relate salt-diffusion rates and ionic current to their driving forces (i.e., salt-concentration and solution-potential gradients, respectively). When normalized by the bulk chemical-diffusivity of salt $D_0$ and bulk ionic-conductivity $\kappa_0$ (i.e., of the pure electrolyte), these tensors exhibit a particular functional form $\underline{\underline{f}}$ (assuming an orthotropic microstructure):

$$\underline{\underline{\kappa}}_{eff}/\kappa_0 = \underline{\underline{D}}_{eff}/D_0 = \underline{\underline{f}} = \varepsilon \begin{bmatrix} \tau_\perp^{-1} & 0 \\ 0 & \tau_\parallel^{-1} \end{bmatrix} \quad [3]$$

Here, the tensor $\underline{\underline{f}}$ depends on local porosity $\varepsilon$ and the local components of tortuosity normal to ($\tau_\perp$) and parallel with ($\tau_\parallel$) the dominant planar orientation. These tortuosity components increase with local microporosity $\varepsilon$ as $\tau_j = \varepsilon^{-\alpha_j}$,[7] where $\alpha_j$ is the tortuosity scaling-exponent for transport in direction $j$. We model the scaling exponents theorized previously[7] for graphite and LiCoO$_2$ electrodes that are listed in Table I. In general the microstructure's orientation can vary in space, and the transport tensor $\underline{\underline{f}}$ would rotate in the fixed Cartesian-frame:

$$\underline{\underline{f}}_{Cart} = \underline{\underline{R}}^T (\underline{\underline{f}}) \underline{\underline{R}}, \quad [4]$$

where $\underline{\underline{f}}_{Cart}$ is the transport tensor in the fixed Cartesian-frame and $\underline{\underline{R}}$ is a space-dependent rotation-matrix describing the microstructure's local orientation. For the two-dimensional system here $\underline{\underline{R}}$ is given in terms of the polar angle $\theta$ between the x-axis and the direction perpendicular to the plane of microstructural orientation [see Fig. 1a]:

$$\underline{\underline{R}} = \begin{bmatrix} \cos\theta & -\sin\theta \\ \sin\theta & \cos\theta \end{bmatrix}, \quad [5]$$

Table I. Pore-space tortuosity scaling-exponents[7] used in the present simulations.

| Exponent | Graphite | LiCoO$_2$ |
|---|---|---|
| perpendicular (thru-plane), $\alpha_\perp$ | 1.914 | 0.830 |
| parallel (in-plane), $\alpha_\parallel$ | 0.600 | 0.640 |

Here, we assume uniform orientation along the plane of the current collector [i.e., $\theta(x, y) = \pi/2$].

The normal $\kappa_{eff,\perp}$ and parallel $\kappa_{eff,\parallel}$ components of the effective ionic-conductivity tensor are shown as a function of macro-pore coverage in Fig. 2b for a pressed-graphite anode with 30% average porosity. Each macro-pore coverage level $\nu_{mp}$ has a corresponding microporosity $\varepsilon$ given by Eq. 1. The normal and parallel components of tortuosity decrease with increasing microporosity. Consequently, both components of effective conductivity (normal and parallel) decrease with increasing macro-pore coverage because microporosity decreases with increasing macro-pore coverage. The anisotropy of effective ion conductivity[7] is also evident from Fig. 2b, where conductivity perpendicular to the current collector $\kappa_{eff,\perp}$ is more than three times lower than the conductivity parallel to the current collector $\kappa_{eff,\parallel}$.

*Governing equations.*— Porous-electrode theory[11,12] is used presently to model the coupling between ion conduction, electron conduction, Li intercalation, and electrochemical reactions in heterogeneous, composite electrodes. Ion conduction occurs in the pores of the heterogeneous electrode. The particular electrolyte considered here contains 1 mol/L LiPF$_6$ salt in mixed-carbonate solvent that exhibits a bulk ionic-conductivity $\kappa_0$ and bulk chemical-diffusivity of salt $D_0$ that varies with the local concentration of salt $c_e$ [see Refs. 13,14 and Figs. 2c,2d]. Ion conduction in the present monovalent, binary electrolyte (with constant transference number) is governed by charge conservation and salt conservation for a concentrated solution:[11,12,15,16]

$$\nabla \cdot \left[ -\underline{\underline{\kappa}}_{eff} \left( \nabla\phi_e - \frac{2R_gT}{F}\gamma_\pm(1-t_+)\nabla\ln c_e \right) \right] - a\nu_s i_n = 0, \quad [6]$$

$$\frac{\partial(\varepsilon c_e)}{\partial t} + \nabla \cdot (-\underline{\underline{D}}_{eff}\nabla c_e) - a\nu_s \frac{i_n}{F}(1-t_+) = 0, \quad [7]$$

where $\phi_e$ is the solution-phase potential and $c_e$ is the salt concentration in the electrolyte. These conservation equations incorporate the effects of ion-transport anisotropy through the effective-transport tensors $\underline{\underline{\kappa}}_{eff}$ and $\underline{\underline{D}}_{eff}$ that were described previously. Here, $R_g$, $F$, and $T$ are the universal gas-constant, Faraday's constant, and temperature, respectively. For the Li-ion transference number $t_+$ we assume a constant value of 0.38,[13] and we assume a constant value of unity for the thermodynamic factor[15] $\gamma_\pm$, which is defined in terms of the mean-molar activity-coefficient $f_\pm$ as $\gamma_\pm = (1 + \partial \ln f_\pm/\partial \ln c_e)$. The source terms in these equations are proportional to both the reaction current-density $i_n$ at the electroactive-particle/electrolyte interface and the surface-area per unit-electrode-volume $a$ of electroactive particles (see Table II). Electron conduction occurs through conductive material driven by the solid-phase potential $\phi_s$ and is governed by:

$$\nabla \cdot (-\sigma_s \nabla\phi_s) + a\nu_s i_n = 0, \quad [8]$$

where $\sigma_s$ is the effective electronic conductivity whose value (see Table II) is taken from the literature.[17]

Table II. Electrochemical-transport material-parameters used in the present simulations.

| Parameter | Graphite | LiCoO$_2$ |
|---|---|---|
| electronic conductivity, $\sigma_s$ (S/m) | 10[†] | 10[†] |
| volumetric surface-area, $a/\nu_s$ (m$^2$/m$^3$) | 3 × 10$^6$ | 3 × 10$^6$ |
| terminal lithium concentration, $c_{s,max}$ (mol/m$^3$) | 30,555[†] | 51,554[†] |
| rate constant, $k$ (mol/m$^2$-s × m$^{4.5}$/mol$^{1.5}$) | 5.03×10$^{-11}$[†] | 2.34×10$^{-11}$[†] |

[†]Ref. 17.

Electrochemical reactions between the electrolyte and solid-state electroactive particles produce gradients of intercalated-Li concentration in electroactive particles. Here, the concentration of intercalated Li $c_s$ is given in terms of the intercalated-Li fraction $x_{Li}$ and the terminal concentration of intercalated Li $c_{s,max}$ as $c_s = x_{Li}c_{s,max}$. For the present electroactive particles (2 μm diameter for both graphite and LiCoO$_2$) at the C-rates of interest (0.5 C to 3 C), intercalated-Li gradients have been neglected in electroactive particles because the difference is less than 0.25% for the cathode and 0.1% for the anode between the intercalated-Li fraction at the surface of and on average in electroactive particles cycled at 3 C.[18] Under these conditions Li conservation in the solid phase is given by:

$$\nu_s c_{s,max} \frac{\partial x_{Li}}{\partial t} + \nu_s a \frac{i_n}{F} = 0, \quad [9]$$

where $\nu_s$ is the volume fraction of the electroactive material.

The driving force for electrochemical reactions is the surface overpotential at the solid/electrolyte interface, $\eta = \phi_s - \phi_e - \phi_{eq}$, where $\phi_{eq}$ is the equilibrium potential of the electroactive material that depends on the fraction of intercalated Li. Functions from the previous literature (Ref. 17) for graphite and (Ref. 19) for LiCoO$_2$ were used to model the variation of equilibrium potential $\phi_{eq}$ versus the fraction of intercalated Li $x_{Li}$ in each electrode. Electrochemical reaction-kinetics are modeled with the Butler-Volmer equation for reaction current-density $i_n$:[16]

$$i_n = i_0 \left[ \exp\left(\frac{0.5F\eta}{R_gT}\right) - \exp\left(-\frac{0.5F\eta}{R_gT}\right) \right]. \quad [10]$$

The exchange current-density $i_0$ depends on the concentration of salt in the electrolyte $c_e$, the fraction of intercalated Li $x_{Li}$, and the kinetic rate-constant $k$ for the particular reaction (see Table II):[20]

$$i_0 = Fkc_{s,max}(c_e)^{0.5}(1-x_{Li})^{0.5}(x_{Li})^{0.5}. \quad [11]$$

*Boundary conditions.*— The average current-density $i$ applied to the cathode's current collector is held constant to simulate galvanostatic charge and discharge process. During galvanostatic cycling the anode current-collector is grounded with solid-phase potential $\phi_{s,-}$ set to 0 V, and the solid-phase potential of the cathode current-collector $\phi_{s,+}$ floats, such that the cell voltage ($\phi_{s,+} - \phi_{s,-}$) varies with time. The flux of Li ions is zero in the outward normal direction at the current collectors. At the separator a null electronic-current boundary-condition is imposed to replicate the electronically insulating property of the separator. Periodic boundary conditions are used to relate the solid-phase potentials, solution-phase potentials, and salt concentration on the opposing sides of the two-dimensional domain. The ionic resistance due to the separator is presently neglected because its thickness (∼10 μm) is small relative to that of the cell sandwich (100–400 μm).

*Numerical discretization, model implementation, and validation.*— The governing equations are discretized using the finite-volume method with implicit differencing and central differencing in time and space, respectively.[21] The fully coupled set of equations is solved iteratively to resolve non-linearities in the governing equations.[19] The solution algorithm is composed of two nested iteration-loops. In the inner loop, non-linearities in the kinetics of intercalation are resolved with respect to lithium concentration in the electroactive material and in the electrolyte using Newton-Raphson iteration and the algebraic-multigrid method.[22–25] If the change in kinetic overpotential between successive iterations exceeds a threshold value (200 mV presently), an under-relaxation factor for the kinetic overpotential (10% presently) is used to stabilize convergence of the iterative sequence until changes in kinetic overpotential decrease below the specified threshold value. In the outer loop, non-linearities in ionic conductivity are resolved with respect to salt concentration in the electrolyte using deferred correction, and no relaxation parameters are used to stabilize these iterations. Convergence of the iterative scheme is achieved when potentials differ by less than $10^{-9}$ V and salt concentrations differ by less than $10^{-9}$ mol/L between values during successive iterations.

For two-dimensional implementations of porous-electrode theory, the interpolation schemes for ion transport parameters and the choices for the maximum time-step can dramatically affect the stability of the numerical scheme and its convergence. We note that the issues described subsequently are not apparent in the one-dimensional implementation of porous-electrode theory that is ubiquitous in the literature. We use harmonic-mean formulas to interpolate effective salt-diffusivity values at the faces of finite-volume cells. In contrast, the bulk ionic-conductivity on a given face of a finite-volume cell is approximated by that of the upwind cell-centroid, where the upwind direction is determined by the direction of the product of (1) the migration velocity $\vec{v}_m$ for Li ions in the electrolyte ($\vec{v}_m = -\underline{\underline{\kappa}}_{eff} \nabla \phi_e / \varepsilon F c_e$) and (2) the slope $\delta_\kappa(c_e)$ of the bulk ionic-conductivity with respect to salt concentration ($\delta_\kappa(c_e) = \partial \kappa_0 / \partial c_e$). Similar schemes have been used previously to discretize the Poisson-Nernst-Planck equations for dilute solutions with the finite-volume method.[26] The latter factor accounts for the non-monotonic variation of bulk ionic conductivity at moderate salt concentrations [see $c_e > 0.5$ mol/L in Fig. 2c].

Furthermore, solution of the discrete equations for ion transport can converge slowly and be altogether unstable when the velocity of Li-ion migration ($\vec{v}_m = -\underline{\underline{\kappa}}_{eff} \nabla \phi_e / \varepsilon F c_e$) exceeds a critical value of $\Delta u/\Delta t$, where $\Delta u$ is the discrete spacing between cell centroids. When cells are cycled at high enough rates the electric field magnitude $|\nabla \phi_e|$ becomes large and salt concentration depletes ($c_e \to 0$ mol/L), causing migration velocity to diverge and exacerbate this stability issue. Similar instabilities have been observed in the numerical solution of the drift/diffusion equations for electron/hole transport in semi-conductors, where Scharfetter and Gummel[27] developed an exponential interpolation scheme to stabilize numerical solutions to these equations. In lieu of using a similar exponential scheme, we stabilize the solution by adaptively reducing the time step such that $\Delta t \leq CFL_{m,\text{tol}} \Delta u / |\vec{v}_m \cdot \hat{n}|$, where $CFL_{m,\text{tol}}$ is a Courant-Friedrichs-Lewy-number tolerance for migration (which we often take as 0.5) and $\hat{n}$ is the unit-normal on the cell-face spaced at $\Delta u$.

The present model was validated with Dualfoil 5.0 (Ref. 11) for a LiCoO$_2$ (50 vol.%)/graphite (60 vol.%) cell with 100-micron-thick homogeneous electrodes, 1 mol/L LiPF$_6$ salt in EC:DMC solvent, and tortuosity-scaling exponents of 0.5. Equilibrium potentials and bulk electrolyte properties were taken as those provided in Dualfoil 5.0 for the materials of interest. The simulated parameters that were compared with the present model were (1) the cell voltage over the discharge process, (2) the salt concentration profile at 30 and 60 minutes, and (3) the solution-phase potential at 30 and 60 minutes. Among all the data compared with the results of Dualfoil 5.0, the difference was less than the numerical precision with which data is output by Dualfoil 5.0.

### Results and Discussion

We first present results that elucidate the ion transport mechanisms that occur in bi-tortuous electrodes. Subsequently, we present a systematic study of discharge capacity for a variety of bi-tortuous structures, identifying key design-parameters and their influence on cycling performance. Lastly sensitivity analysis of optimized bi-tortuous electrodes designs is conducted to quantify how the performance of these electrodes is affected by cycling rate, electrode thickness, average loading, and average porosity. The metric used to quantify performance is discharge capacity, defined as the percentage of charge transferred during the discharge process relative to the theoretical maximum under open-circuit conditions.

The results presented in the first sub-section are for full-cells cycled at a C-rate of C/2 having 200 μm-thick electrodes with 30% average porosity and 70% average volume-fraction of electroactive material. In the second sub-section these parameters are varied for optimized bi-tortuous anode designs. In all cases we benchmark the performance of bi-tortuous anodes with homogeneous anodes having the same average porosity (and loading) and electrode thickness. In all cases the same values of electrode thickness, average porosity, and electroactive-material volume-fraction are used for both the cathode and anode. We also assume that all space not filled by electrolyte is occupied by electroactive-material (i.e., neglecting the volume of binder and conductive carbon).

*Ion transport during galvanostatic cycling.*— Figure 3 shows the development of voltage with time for the charge/discharge process at C/2 for three different cases: (a) a homogeneous anode containing no macro-pores, (b) a bi-tortuous anode containing macro-pores having 20% coverage and spaced at wide intervals ($s/w = 2.0$), and (c) a bi-tortuous anode containing macro-pores having 20% coverage and spaced at short intervals ($s/w = 0.5$). The thumbnails inside the voltage plots correspond to the anodic structure for each of the cases and the snapshots adjacent to the plot represent the fraction of intercalated Li at five instants in time indicated on the voltage-versus-time curves in Fig. 3. At two instants in time (2 and 3) the lines along which ionic current density flows (i.e., the ionic current-density lines) are shown, where the ionic current-density vector $\vec{i}_e$ at a given point in space is defined as $\vec{i}_e = -\underline{\underline{\kappa}}_{eff} (\nabla \phi_e - 2\gamma_\pm (1-t_+) RT/F \nabla \ln c_e)$.

For each voltage-versus-time curve the cell voltage increases from 3 V to 4 V (cutoff) during the charging process. Immediately afterward the discharge process starts and cell voltage decreases as time proceeds. Charge/discharge curves are shown as loops in order to illustrate the effect of polarization, which is proportional to the difference in voltage between charge and discharge. Comparing each of the voltage curves with the theoretical limit (which is 120 min at C/2) it is evident that the battery with a homogeneous anode has poor performance [Fig. 3a], while a cell having optimized macro-pores reaches a capacity much closer to the theoretical limit [Fig. 3c]. Also the bi-tortuous electrode in Fig. 3c shows smaller polarization than the homogeneous electrode. The mechanisms explaining the improved performance of the design in Fig. 3c and lower performance of the design in Fig. 3b are explained below.

As stated previously, the high anisotropy of pressed-graphite electrodes leads to low ion-conductivity (i.e., high tortuosity) in the direction perpendicular to the current collector. As a result, a large potential difference is required to conduct Li ions through the anode's depth. However since the cell's charging voltage is limited to 4 V, Li intercalates to a small degree throughout much of the anode's depth (see

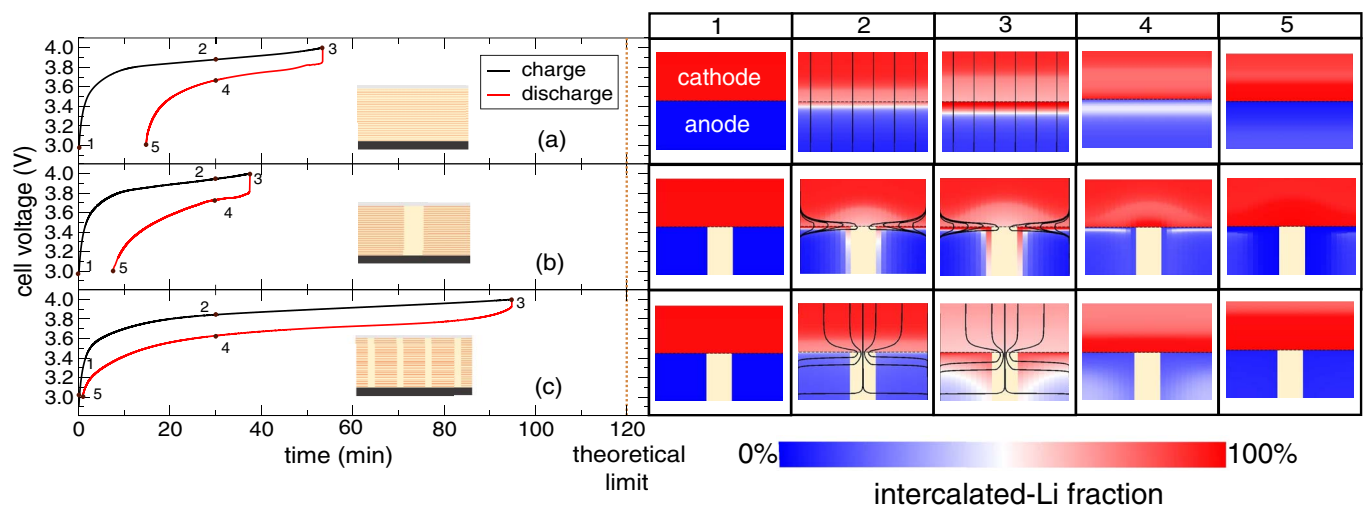

**Figure 3.** Voltage as a function of time for galvanostatic charge and discharge at C/2 rate for the following cases: (a) a homogeneous anode, (b) a bi-tortuous anode with $s/w = 2.0$ and 20% macro-pore coverage, and (c) a bi-tortuous anode with $s/w = 0.5$ and 20% macro-pore coverage. Snapshots of intercalated-Li fraction (not drawn to scale) at different instants in time marked on the voltage curves are shown as well. Lines of ionic current-density are shown at two instants in time for the three cases.

Fig. 3a, snapshot 3). In contrast, deintercalation of Li in the cathode (LiCoO$_2$) proceeds over a larger extent than in the anode, because the LiCoO$_2$ cathode is more isotropic and has a higher thru-plane ionic conductivity than the graphite anode.

For cases 2 and 3 that contain macro-pores, Li intercalation initiates along the macro-pore's edge (Figs. 3b,3c, snapshot 3). This initial reaction front becomes curved as it propagates into the interior of the anode. However in case 2 the distance between macro pores $s$ is large and ion transport parallel to the current collector is impeded by this large distance. Also, incorporation of a macro-pore with 20% coverage reduces the local porosity from 30% (for a homogeneous electrode) to 12.5% for the same average porosity. This produces an additional increase in the transverse ionic resistance inside the anode. As a result of these two effects, the reaction front in the anode becomes pinned near the separator at the end of charging cycle (Fig. 3b, snapshot 3). We note that intercalation proceeds along the edges of the macro-pore but only within a small region near the macro-pore's surface. Similarly, at the end of discharge (Fig. 3b, snapshot 5), capacity is underutilized in the regions between macro-pores in the anode.

Case 3 exhibits much better cycling performance than case 2, because of the shorter distance between macro-pores. The proper balance between the macro-pore coverage (effecting local micro-porosity) and the spacing between macro-pores enables high utilization of graphite (as much as 80%). This is apparent from the high intercalated-Li fraction throughout the entire anode at the end of charging process (Fig. 3c, snapshot 3).

In order to further elucidate this, lines of ion current-density are plotted for the three cases. The homogeneous electrode shows ion current-density lines that are perpendicular to the current collector (Fig. 3a, snapshots 2, 3). When the spacing between macro-pores is large [Fig. 3b] the intercalation of Li is concentrated close to the separator and at the interface between the macro-pore and graphite. As mentioned above, the large distance between macro-pores impedes transverse ion transport; consequently, ion current-density lines concentrate around the separators (Fig. 3b, snapshots 2, 3). When the spacing-to-thickness ratio is small [Fig. 3c], ion current-density lines route preferentially through the macro-pore and extend transversely into the micro-porous region of the anode. The intercalation of Li occurs through the depth of anode (as does the delithiation in cathode), and it is evident that the primary direction of ion transport in the porous, graphite anode is parallel to the current collector.

*Optimizing macro-pore design: coverage, spacing, shape, and placement.—* The results in the previous section demonstrate that the dimensions and arrangement of macro-pores dictate their effectiveness in enhancing cycling performance. Here, effects of macro-pore coverage, spacing, shape, and placement are explored systematically. Firstly, the effect of macro-pore coverage and shape is studied. Figure 4 shows the discharge capacity obtained for cycling LiCoO$_2$/graphite cells having particular macro-pore designs in the anode of a given cell. Straight macro-pores having certain coverage and spacing levels are considered first, though tapered shapes are considered later.

For a spacing-to-thickness ratio of $s/w = 0.12$ discharge capacity increases monotonically with increasing macro-pore coverage. For larger spacing-to-thickness ratios (e.g., $s/w = 0.50$) discharge capacity is maximized at a particular macro-pore coverage level. This optimum appears as macro-pore coverage increases because of the competition between (1) increasing in-plane transport resistance in the micro-porous region of the anode and (2) decreasing thru-plane resistance of the macro-pore. Initially, increasing macro-pore coverage produces an increase of discharge capacity because of reduced thru-plane resistance to ion transport. However beyond the optimal coverage level, local porosity in the micro-porous region of the anode approaches zero [see Fig. 2a] leading to high resistance in both the in-plane and thru-plane directions. As a consequence of the

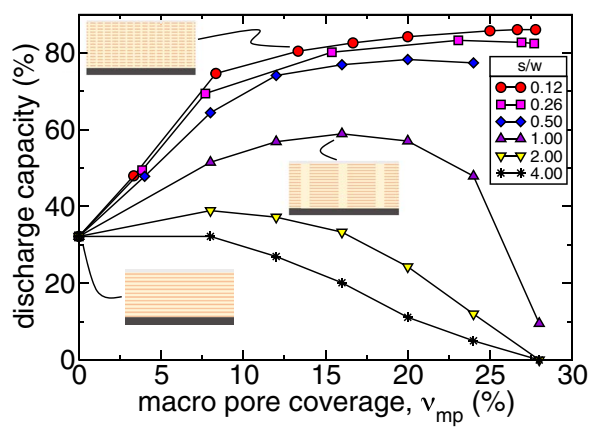

**Figure 4.** Variation of discharge capacity with macro-pore coverage for various spacing-to-thickness ratios in a graphite anode with 30% average porosity, 200 μm thickness, and cycled at C/2 rate.

A1420	Journal of The Electrochemical Society, 162 (8) A1415-A1423 (2015)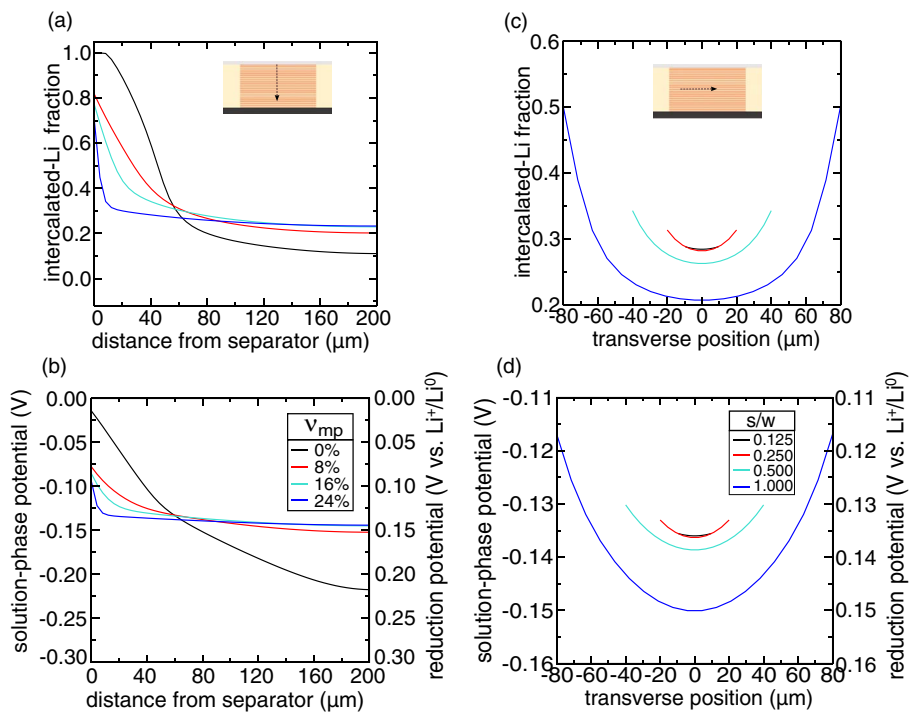

**Figure 5.** Profiles of (a) intercalated-Li fraction and (b) solution-phase potential through the thickness of graphite anodes (as shown in the thumbnail). These profiles are shown for increasing macro-pore coverage with a particular spacing-to-thickness ratio of 0.5. Profiles of (c) intercalated-Li fraction and (d) solution-phase potential across the transverse direction and through the center of graphite anodes. These profiles are shown for increasing spacing-to-thickness ratio and having fixed macro-pore coverage of 25%. All profiles are shown after 50 min from the start of the charging process at C/2.

polarization induced by ion-transport resistance, charge/discharge capacity reduces below the theoretical limit.

Incorporating a macro-pore into a graphite anode will not always result in a cell with enhanced performance relative to a homogeneous electrode. For instance, increasing the spacing between macro-pores beyond a certain limit at a fixed coverage level (e.g., $s/w \geq 2$ for 15% coverage) can result in poorer performance than a homogeneous electrode of the same average porosity (see Fig. 4, yellow triangles). Also, as the spacing-to-thickness ratio increases, the maximum discharge-capacity decreases. The optimal macro-pore coverage at which the maximum discharge-capacity is obtained shifts to low macro-pore coverage levels as the spacing-to-thickness ratio increases. This result reveals that, in general, the effectiveness of macro-pores decreases as spacing-to-thickness ratio increases. In other words, macro-pore dimensions must be sufficiently small if they are going to produce a measureable enhancement in cycling performance.

To understand the particular mechanisms that produce the behavior displayed in Fig. 4, solution-phase potential and intercalated-Li profiles are plotted in Fig. 5. In Figs. 5a,5b these profiles are shown through the thickness of the graphite anode, and the curves shown correspond to various macro-pore coverage levels, all after 50 minutes of charging each cell and with the same spacing-to-thickness ratio ($s/w = 0.5$). The magnitude of the potential drop across the graphite anode decreases as macro-pore coverage increases [Fig. 5b]. This trend occurs because more current flows through macro-pores as coverage increases. Consequently, less current flows through the thickness of the micro-porous region and potential drop is less across it as a result. We also note that the large potential drop in the homogeneous anode is coincident with the termination of charging and that complete intercalation of graphite near the separator occurs as well [Fig. 5a]. In contrast, cells with bi-tortuous anodes show less extreme variation of intercalated-Li fraction. The plateau of intercalated-Li fraction far from the separator is a result of uniform reactions that occur inside the bi-tortuous anode. In reality, cells cycled to extreme states-of-charge would be accompanied by the deposition of Li metal that would induce rapid capacity fade. This tendency is also reflected by the reduction potential (defined locally inside the electrode as $\phi_s - \phi_e$) approaching 0 V vs. Li$^+$/Li$^0$ near the separator [Fig. 5b]. Thus, macro-pores could improve capacity retention in LiCoO$_2$/graphite cells, in addition to enhancing their discharge capacity, since extreme states-of-charge can be avoided. Previous work[29] has predicted that Li-metal deposition is exacerbated at electrode edges, but extension of anode surfaces beyond the cathode helped to prevent such deposition. The results shown in Fig. 5d reveal that variations in reduction potential are most extreme for macro-pores with large spacing. These trends suggest that Li-metal deposition could be a problem for macro-pores spaced at large intervals, but Li-metal deposition is unlikely for designs with sufficiently small spacing (e.g., the 100 μm spacing case shown in Fig. 5b). The effect of macro-pore spacing is apparent from the solution-phase potential-drop (which produces a variation in reduction potential) which is smallest for short spacings. Solution-phase potential and intercalated-Li fraction profiles are also plotted in the in-plane direction through the center of the graphite anode [Fig. 5c,5d]. The curves shown correspond to various spacing-to-thickness ratios, all after 50 minutes of charging each cell and with 15% macro-pore coverage. More frequently spaced macro-pores (given by small $s/w$ values) produce more uniform Li intercalation in the transverse direction. Li intercalation becomes more uniform across the anode when spacing between macro-pores decreases, because in-plane resistance of the micro-porous region of the electrode is reduced. This effect is reflected by the reduction in solution-phase potential-drop across the anode when spacing is decreased [Fig. 5d].

A straight macro-pore shape is not necessarily optimal electrochemically and may be difficult to manufacture, since a line-of-sight process would be required to produce it. To quantify the effect of pore shape on electrochemical performance, we have also considered an angled taper on the pore's sidewall. A single macro-pore of this type was simulated in Ref. 10. Fig. 6 shows the charge capacity of such a cell as a function of taper angle for a fixed macro-pore coverage of 24% and spacing-to-thickness ratio of $s/w = 0.5$ (for an electrode thickness of 200 μm). For this particular coverage and spacing, the maximum taper-angle was 6.84° for a macro-pore extending completely through the anode's thickness. Positive taper-angles enhance cycling capacity by ∼5% over that of the straight macro-pore, while negative taper-angles produce a dramatic drop in capacity (>50%). This dependence on taper angle is a result of the fact that the amount of thru-plane ionic-current flowing from the cathode to the anode is largest near the separator. Negative taper angles constrict ionic current near the separator to a small cross-sectional area. This constriction results in high current-density and large potential drop through the macro-pore. In the opposite sense, positive taper angles reduce the thru-plane current-density flowing through the macro-pore, reducing solution-phase potential-drop through the macro-pore.

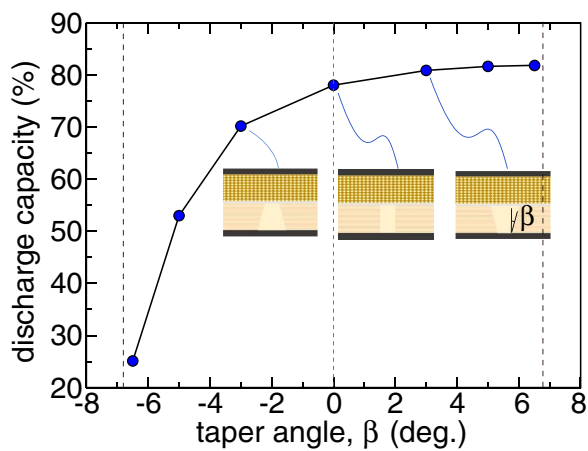

**Figure 6.** Discharge capacity as a function of macro-pore taper-angle. Macro-pore coverage in the graphite anode is fixed to 24% in all cases with 30% average porosity, spacing-to-thickness ratio of 0.5, and cycled at C/2.

The previous results demonstrate that macro-pores in anisotropic graphite anodes can enhance full-cell cycling performance, but we find here that performance can be further improved by structuring the $LiCoO_2$ cathode together with the graphite anode. This improvement is shown in Fig. 7, where discharge capacity is shown as a function of macro-pore coverage in the cathode with 15% macro-pore coverage in the anode and $s/w = 0.5$ for macro-pores in both electrodes (red squares). Here, the discharge capacity increases with increasing macro-pore coverage to a maximum discharge capacity at 25% cathode macro-pore coverage, after which discharge capacity decreases. As a reference, discharge capacity is shown for various macro-pore coverage-levels in the anode alone (i.e., without macro-pores in the cathode). A clear trend of improved performance is observed: (1) the homogeneous electrode achieves 35% discharge capacity, (2) the anode with optimized macro-pore coverage and spacing achieves 80% discharge capacity, and (3) the cell with both electrodes having optimized macro-pores achieves 90% discharge capacity. Using macro-pores in the cathode cuts the amount of capacity lost in half relative to the case with macro-pores in the anode alone. Further, the present analysis assumes that the cathode macro-pores align with anode macro-pores, which provides an upper bound on the enhancement in capacity that could be expected if macro-pores were not aligned in the respective electrodes. In practice, lateral alignment between the two electrodes during cell assembly will be difficult because of the fine features that the electrodes possess (∼10 μm).

*Sensitivity of performance: Cycling rate, electrode thickness, average porosity, and electroactive-material loading.—* The results presented thus far are for full-cells cycled at a C-rate of C/2 and having 200 μm-thick electrodes with 30% average porosity and 70% average volume-fraction of electroactive material. Optimized cells may have different electrode thicknesses, average porosities, and electroactive-material loading depending on their particular material, manufacturing, and cost constraints. Further, the rate at which a given cell is cycled will vary during its use and will depend on the particular application in which the cell is used. Thus, it is important to understand how the performance of optimized bi-tortuous electrode structures is affected by cell-manufacturing and operating parameters. Subsequently, we predict the sensitivities of performance of optimized bi-tortuous anodes with respect to these parameters.

Firstly, we consider the sensitivities of performance with respect to cycling rate and electrode thickness. Figure 8 shows discharge capacity as a function of C-rate for cells having three different electrode thicknesses (200 μm, 100 μm, and 50 μm). Figure 8a shows that thicker electrodes have lower capacity than thinner ones at a given C-rate. Also, for a given electrode thickness the bi-tortuous electrode produces higher capacity than the homogeneous electrode at all C-rates. Figure 8c shows the difference in capacity between bi-tortuous and homogeneous electrodes. The maximum enhancement in capacity for the bi-tortuous electrode is obtained at a particular C-rate that increases as electrode thickness decreases.

The sensitivity with respect to electrode thickness and rate can be simplified by considering the theoretical scaling of polarization among these cases. A simple equivalent-circuit model of ohmic, ionic conduction within these electrodes suggests that cell polarization scales proportionally with average applied current-density $i$ and electrode thickness $w$. Figure 8b shows the capacity of these cells as a function of $iw$ (the product of the average current-density and electrode thickness). For sufficiently large electrode thicknesses ($w \geq 100$ μm) these curves collapse on each other, because ohmic, ionic conduction dominates polarization for large enough electrodes. For thinner electrodes (e.g., 50 μm) capacities of both bi-tortuous and homogeneous anodes are less than those of thicker electrodes at a given value of $iw$, because other mechanisms than ionic conduction (e.g., electrochemical reaction-kinetics and solid-state mass-transfer) contribute more cell polarization for thin electrodes.

The average porosity and loading of electroactive material are important to study from the standpoint of manufacturing constraints and cost: (1) porosities will be limited by the packing density of electroactive-material (see Ref. 6) and the amount inactive additives used (e.g., binder and conductive carbon) and (2) the average porosity will affect the total amount of electrolyte used for a given cell and will affect the cost-contribution from it (relative to the total cell-cost). Accordingly, capacity is shown in Fig. 9 as a function of average porosity (and average loading of electroactive material) for anodes with various macro-pore coverage-levels. In all cases we fix $s/w = 0.5$ and $w = 200$ μm. Here, the particular C-rate (or average current density) used to cycle a cell of certain average porosity was varied proportionally with the average volume-fraction of active material $\bar{v}_s$ and the thru-plane effective ionic conductivity of the benchmark homogeneous-electrode $\kappa_{eff,\perp} \propto \bar{\varepsilon}^{2.914}$ at the average porosity $\bar{\varepsilon}$ [see Fig. 9b]. For these C-rates, bi-tortuous structures having 20% and 40% macro-pore coverage achieve discharge capacities of 80–90% for average porosities ranging between 20–80% [Fig. 9a]. In all cases, the bi-tortuous structures show higher discharge capacity than their homogeneous counterparts having the same average porosity and loading [Fig. 9a]. We note that the enhancement gained by the bi-tortuous structures (relative to their homogeneous counterparts) decreases as porosity increases. Therefore, the present bi-tortuous structures will be particularly effective when minimal amounts of electrolyte are used inside the battery.

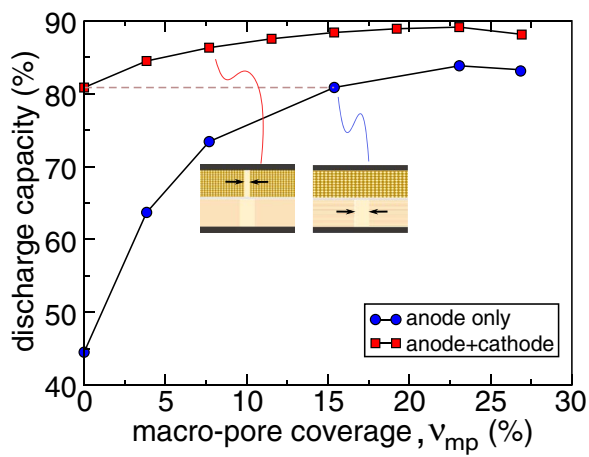

**Figure 7.** Discharge capacity as a function of cathode macro-pore coverage with fixed anode macro-pore coverage of 15%. As a benchmark, discharge capacity is plotted as a function of anode macro-pore coverage for a cell having no macro-pore in the cathode. Both electrodes have an average porosity of 30% and spacing-to-thickness ratio of 0.5. Discharge capacity was obtained by cycling at C/2 rate.

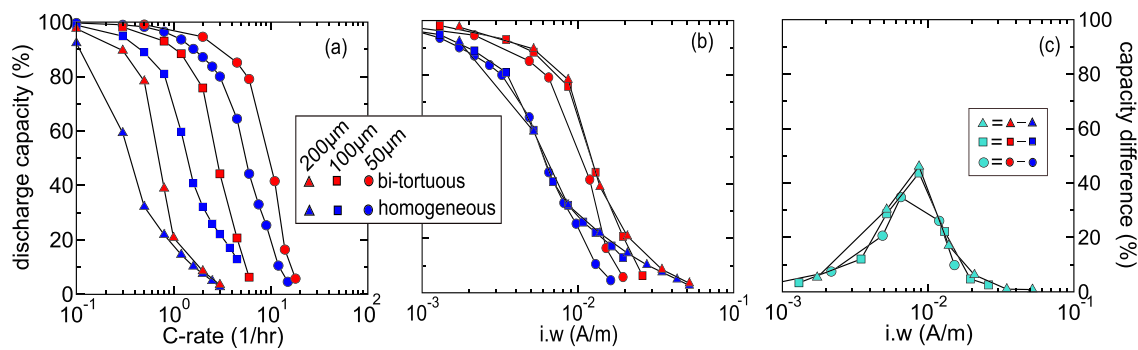

**Figure 8.** (a) Discharge capacity as a function of C-rate. Six cases are shown: three different electrode thicknesses, each with a bi-tortuous (20% macro-pore coverage and spacing-to-thickness ratio of 0.5) or homogenous anode. In (b) discharge capacity is plotted against the product of average applied current-density $i$ and the electrode thickness $w$. (c) The difference in capacity between the bi-tortuous and homogeneous electrodes is shown as a function of $iw$.

## Conclusions

We have predicted that bi-tortuous electrode structures in anisotropic graphite anodes can enhance discharge capacity by two-fold over a homogeneous electrode containing the same average porosity. The particular design of macro-pores in a given bi-tortuous structure impacts the realizable degree of enhancement, and macro-pores must be sized optimally and spaced at short intervals to realize the maximum enhancement. In practice, manufacturing constraints will also influence the types of macro-pore shapes that are feasible. We note that this is an open area of research to which our results can be applied to determine trade-offs between electrochemical performance and manufacturability. In addition, the particular graphite-orientation that we have considered could be different (depending on manufacturing conditions), but spatial orientation-variations can be incorporated in our formalism through the effective transport-tensors.

In a more general sense, our approach provides additional parameters with which cell performance can be controlled. For example, typically the geometry of an electrode with a given active material is defined with just two parameters that can be controlled in the manufacturing process, thickness and porosity. Our approach adds a number of additional parameters describing macro-pores that can potentially be controlled in the manufacturing process, allowing simultaneous optimization with respect to more constraints (including cost, power, energy, and volume).

While we have not modeled Li-metal deposition explicitly in this work, our findings suggest that macro-pores could also limit the capacity fade from such a process by reducing variability in the fraction of intercalated-Li inside of a graphite anode. Further, we also show that placement of a complementary macro-pore in the cathode provides a mild enhancement in discharge capacity. The performance of bi-tortuous graphite anodes is also shown to depend on the cycling rate. Our simulations suggest that bi-tortuous anodes are particularly useful at high rates, but more detailed pulsed-power analysis (rather than continuous, galvanostatic discharge) would be required to evaluate their fitness for plug-in hybrid or electric vehicle applications (such as was done in Ref. 28). Furthermore, we find that bi-tortuous anodes are particularly useful when average porosity is low and average electroactive-material loading is high. In practice, both cost of the cell and the particle-packing constraints introduced by the various constituents inside the composite electrode will dictate what porosity and loading levels are accessible and appropriate. Our future efforts aim toward fabrication of bi-tortuous electrodes, testing of their electrochemical cycling performance, and comparison with the present predictions.

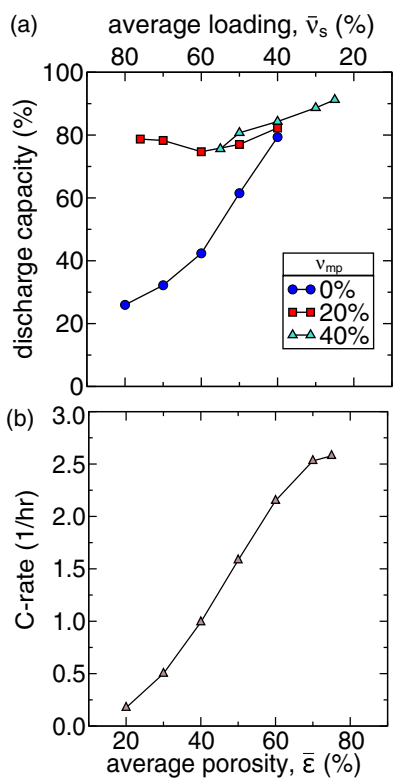

**Figure 9.** (a) Discharge capacity as a function of average porosity (and average electroactive-material volume-fraction/loading) for various macro-pore coverage levels in a graphite anode. Spacing-to-thickness ratio was fixed to 0.5 in all cases. For each porosity the C-rate was chosen as shown in (b).

## Acknowledgments

VPN and KCS thank the Department of Mechanical Science and Engineering at the University of Illinois at Urbana-Champaign for financial support. SJH was supported by the Assistant Secretary for Energy Efficiency and Renewable Energy, Office of Vehicle Technologies of the U.S. Department of Energy under Contract No. DE-AC02-05CH11231, under the Batteries for Advanced Transportation Technologies (BATT) Program. KCS designed the present computational study and VPN simulated electrochemistry in the bi-tortuous structures presented here. KCS developed the two-dimensional porous-electrode theory and implementation in MATLAB. VPN and KCS wrote the first draft and edited the manuscript. SJH conceived of using macro-pores to enhance ion transport in anisotropic graphite anodes and edited the manuscript.

### List of Symbols

| Symbol | Description |
|---|---|
| $a$ | volumetric surface-area, m$^2$/m$^3$ |
| $\alpha_\square$ | tortuosity scaling-exponent of direction $j$, - |
| $\beta$ | taper angle, degrees |
| $C_e$ | salt concentration, mol/m$^3$ |
| $C_{max,s}$ | concentration of inserted Li at saturation, mol/m$^3$ |
| $D_0$ | bulk chemical diffusivity, m$^2$/s |
| $\underline{\underline{D}}_{eff}$ | effective chemical diffusivity tensor, m$^2$/s |
| $\varepsilon$ | porosity, - |
| $\bar{\varepsilon}$ | average porosity, - |
| $f_\pm$ | mean-molar activity-coefficient of salt, - |
| $F$ | Faraday's constant, C/mol |
| $\gamma_\pm$ | thermodynamic factor, - |
| $\eta$ | overpotential, V |
| $i_0$ | exchange current-density, A/m$^2$ |
| $i_n$ | local reaction current-density, A/m$^2$ |
| $\kappa_o$ | bulk ionic-conductivity, S/m |
| $\underline{\underline{\kappa}}_0$ | effective ionic-conductivity tensor, S/m |
| $\bar{\nu}_s$ | average volume-fraction of electroactive material, - |
| $\nu_s$ | volume fraction of electroactive material, - |
| $\nu_{mp}$ | macro-pore coverage, - |
| $\phi_e$ | solution-phase potential, V |
| $\phi_{eq}$ | equilibrium potential, V |
| $\phi_s$ | solid-phase potential, V |
| $R_g$ | universal gas constant, J/mol-K |
| $\sigma_s$ | effective electronic-conductivity, S/m |
| $\theta$ | polar angle, degrees |
| $t$ | time, s |
| $t_+$ | transference number of Li ion, - |
| $T$ | temperature, K |
| $w$ | electrode thickness, m |
| $x$ | thru-plane coordinate, m |
| $x_{Li}$ | intercalated-Li fraction, - |
| $y$ | in-plane coordinate, m |